\documentstyle[aps,prb,epsf,multicol]{revtex}
\tighten
\begin{document}
\draft
\title{
       Frequency-Dependent Shot Noise as a Probe of Electron-Electron 
       Interaction in Mesoscopic Diffusive Contacts
       } 
\author{K.\ E.\ Nagaev}

\address{Institute of Radioengineering and Electronics, Russian Academy
of Sciences, Mokhovaya ulica 11, 103907 Moscow, Russia
\footnote{e-mail address: nag@mail.cplire.ru}
}
\maketitle
\bigskip
\begin{abstract}
The frequency-dependent shot noise in long and narrow mesoscopic
diffusive contacts is numerically calculated. The case of arbitrarily
strong electron-electron scattering and zero temperature of
electrodes is considered. For all voltages, the noise increases with
frequency and tends to finite values. These limiting values are larger
than the Poissonian noise and increase with voltage nearly as
$V^{4/3}$. This allows one to experimentally determine the
parameters of electron-electron interaction.
\end{abstract}
\pacs{Pacs numbers: 72.70.+m, 73.23.-b, 73.50.Td, 73.23.Ps}

\begin{multicols}{2}
\narrowtext

Basically, electron-electron scattering appears in two different
forms. First, this is the decay of quasiparticle states (phase-breaking)
and second, this is relaxation of nonequlibrium distribution function
of electrons. In last two decades, the equilibrium phase-breaking
processes were extensively investigated in relation with
weak-localization corrections to conductivity
\cite{AA85} and universal conductance fluctuations.\cite{AKh85,Lee87}
In particular, it was found \cite{AA85} that in dirty metals the
disorder strongly enhances electron-electron (e-e) interactions at
low temperatures, which even makes questionable the validity of
Landau concept of quasiparticles \cite{Abrikosov} in low-dimensional
systems (for a recent review of e-e scattering in mesoscopic systems
see paper by Blanter \cite{Blanter}). Generally, the theoretical
results are in a reasonable agreement with the experimental data, and
this allows determining the phase-breaking time from
weak-localization experiments \cite{Gershenson}.

By far less is known about the kinetics of electron systems in the
presence of e-e scattering, which results in relaxation of
nonequilibrium distribution of electrons to the Fermian one. While
conserving the total energy of electron system, it smooths down the
peculiarities of distribution function of electrons. For example, it
affects the shape of electron distribution function in a wire placed
between two different reservoir electrodes with a finite voltage drop
between them. Recently, Pothier {\it et al.}\ \cite{Pothier}
performed direct measurements of electron distribution function using
tunnel superconducting probes and determined the parameters of e-e
interaction from their data. However their results appeared to be
inconsistent with any of existing theories of e-e scattering. 

Although the shape of electron distribution function does not affect
the conductivity of metals (except for small quantum corrections), it
is crucial in the semiclassical theory of nonequilibrium noise in
solids.\cite{book} In particular, it enters into the semiclassical
expression for the shot noise in diffusive mesoscopic
contacts.\cite{Nag92} Hence this noise may be used for determining
the parameters of e-e scattering.

The effects of strong electron-electron scattering on the
zero-frequency shot noise were studied within
the electron effective temperature approximation
\cite{Nag95,Kozub95} and were shown to increase the ratio $S_I/2eI$ 
from $1/3$ to $\sqrt{3}/4$. Though this increase was experimentally
observed by Steinbach {\it et al.},\cite{Steinbach} it is difficult
to quantitatively estimate the parameters of e-e scattering from it.
The reason is that the zero-frequency noise is determined by the
shape of distribution function near the middle of the
contact, which becomes Fermian at relatively weak e-e scattering and
does not change with its further increase.

The situation is different for the finite-frequency noise in long and
narrow contacts with strong external screening, i. e., with a close
ground plane \cite{Y97} or coaxial grounded shielding.\cite{Nag98} It
was shown that for noninteracting electrons, the high-frequency shot
noise tends to $eI$ if pile-up of charge in the contact is forbidden
\cite{Y97} and to $2eI$ if it is allowed.\cite{Nag98} More recently, 
Naveh {\it et al.}\cite{Y98} obtained that for strong e-e scattering
and zero temperature of electrodes, the noise infinitely grows with
frequency while remaining linear in current. Below we show that this
is not the case: instead of diverging, the actual high-frequency
noise tends to a finite value nonlinearly depending on voltage. We
also propose a method for quantitative determination of the
parameters of e-e scattering from measurements of zero-temperature
high-frequency shot noise.

\begin{figure}
\epsfxsize15cm
\centerline{\epsffile{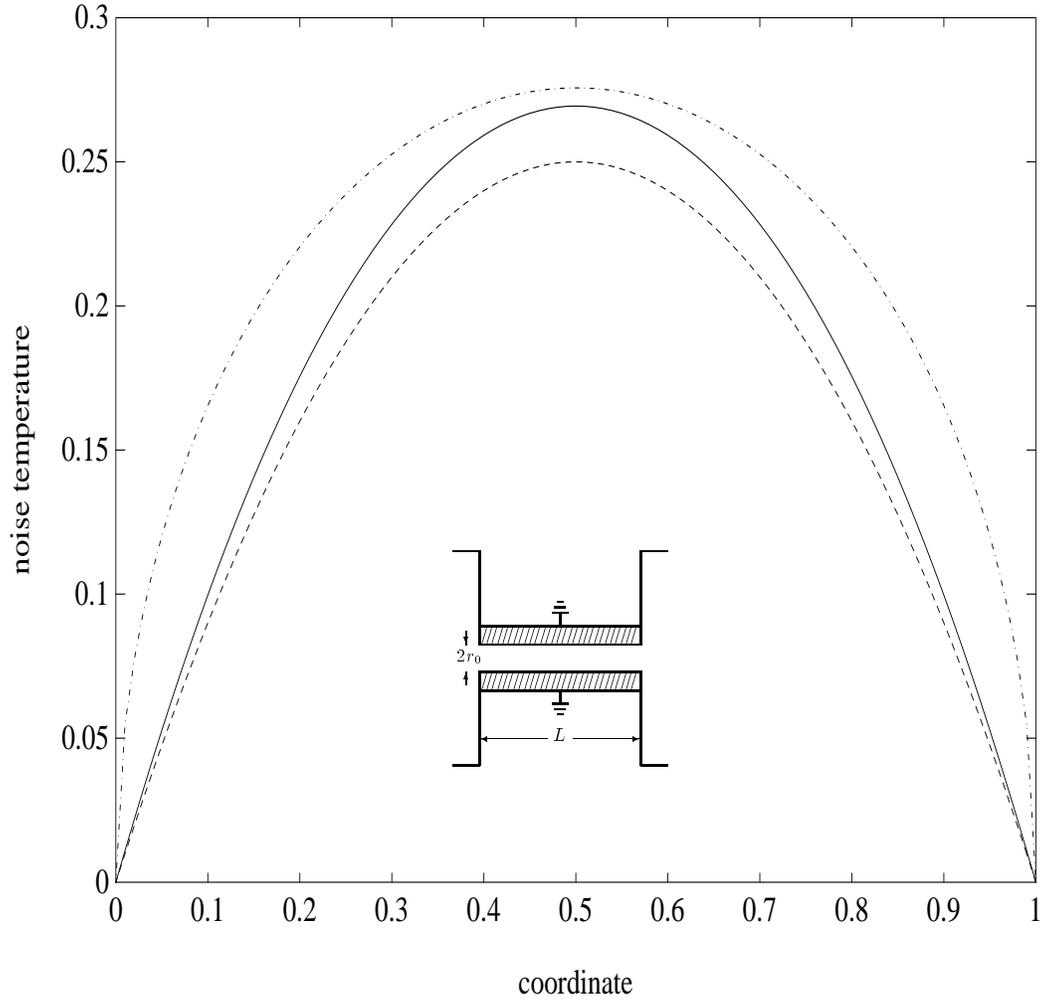}}
\vspace{0.3cm}
\caption
{ 
Dependence of noise temperature $T_N/eV$ on coordinate $x/L$ for
$\alpha = 10$.  The dashed line shows $T_N$ for noninteracting
electrons, and the dash-dot line shows $T_e$ calculated from the
energy-balance equation. Inset shows the longitudinal section of
the shielded contact.
}
\label{temperature}
\end{figure}

In what follows we use the coaxial model,\cite{Nag98} although all
the results also apply to the ground-plane one. Assume that the
contact of length $L$ is a cylinder of circular section with a
diameter $2r_0 \ll L$ consisting of a dirty metal with conductivity
$\sigma$ (see Fig.\ \ref{temperature}, the inset) and connects two
massive electrodes. The contact is screened from the ambient space by
the third perfectly conducting coaxial grounded electrode, which is
separated from its surface by a thin insulating film of thickness
$\delta_0$ and the dielectric constant $\varepsilon_d$. The external
circuit is assumed to have a large grounding capacity, which lifts
the condition of zero net charge of the contact at finite
frequencies.

The spectral density of current noise, e.g., at the
left edge of the contact is given by the formula \cite{Nag98}

\begin{equation}
  S_I( \omega )
  =
  \frac{ 4 }{ RL }
  \int\limits_0^L dx\,
  K( x, \omega ) T_N( x ),
  \label{general}
\end{equation}
where $R$ is the resistance of the contact, $x$ is the longitudinal
coordinate, and $T_N(x)$ is the local ``noise temperature''
determined in terms of the electron distribution function 
$f( \epsilon, x )$ as follows:
\begin{equation}
  T_N(x)
  =
  \int d\epsilon\,
  f( \epsilon, x )
  [ 1 - f( \epsilon, x ) ]
  \label{noise temperature}
\end{equation}
The kernel of the integral (\ref{general}) is given by the
expression 
\begin{eqnarray}
  K( x, \omega )
&
  =
&
  2 ( \gamma_{\omega} L )^2
\nonumber\\
&
  \times
&
  \frac{
        \cosh [ 2\gamma_{\omega} ( L - x ) ]
        +
        \cos [ 2\gamma_{\omega} ( L - x ) ]
       }{
        \cosh ( 2\gamma_{\omega} L  )
        -
         \cos ( 2\gamma_{\omega} L  )
       },
  \label{noise-kernel}
\end{eqnarray}
where  $\gamma_{\omega} = (\omega\varepsilon_d 
/ 4\pi\sigma\delta_0 r_0)^{1/2}$. At sufficiently high frequencies,
the kernel $K$ exponentially decreases with $x$. This decrease has a
simple physical explanation. At contact dimensions much larger than
the Debye screening length, the local current fluctuations inside the
contact, which result from the randomness of impurity scattering,
induce the current fluctuations at the contact edges through
the long-range fluctuations of electrical field. However at finite
frequencies, the electric lines of force emerging from the middle
points of the contact are intercepted by the screening electrode and
do not reach the contact edges. Hence it is only the portions of the
contact adjacent to its edges that contribute to the measurable
noise. Therefore for calculating the high-frequency noise in such
contacts, it is very important to know the exact distribution
function of electrons near their edges.

In our semiclassical approach, the distribution function $f$ obeys
the diffusion equation
\begin{equation}
  D
  \frac{ d^2 }{ dx^2 }
  f( \epsilon, x )
  +
  I_{ee}( \epsilon, x )
  =
  0,
  \label{diffusion}
\end{equation}
where $D$ is the diffusion coefficient. At zero temperature, the
boundary conditions for this equation at the left and the right ends
of the contact are
\begin{equation}
  f( \epsilon, 0 ) = \theta( - \epsilon ),
  \quad
  f(\epsilon, L ) = \theta( eV - \epsilon),
  \label{boundaries}
\end{equation}
where $\theta$ is the step function and $V$ is the voltage drop
across the contact.

Recently, Eq.  (\ref{diffusion}) was solved using the
phenomenological approximation of effective electron temperature,
\cite{Nag95,Kozub95} i.e. the distribution function was
saught in the form
\begin{equation}
  f_T( \epsilon, x )
  =
  \left[
          1
          +
          \exp\left(
                    \frac{ \epsilon - eVx/L }{ T_e(x) }
               \right)
  \right]^{ -1 },
  \label{effective}
\end{equation}
where $T_e$ was the coordinate-dependent temperature of electron
gas. As the collision integral is zero for arbitrary $f_T$ chosen in
form (\ref{effective}),
Eq. (\ref{diffusion}) reduces to an energy-balance equation, whose
solution at zero temperature of electrodes is
\begin{equation}
  T_e
  =
  eV 
  \sqrt{ 3x ( L - x ) }
  /
  \pi L.
  \label{root}
\end{equation}
As $T_e$ exhibits squire-root singularities at the edges of the
contact, substitution of Eq. (\ref{root}) into (\ref{general})
results in high-frequency noise diverging \cite{Y98} as
$\omega^{1/4}$. This unphysical divergency is due to the inadequate
description of the distribution function near the contact edges by
the effective temperature model. Indeed, for $f = f_T$ with $T_e$
given by Eq. (\ref{root}), the first term of Eq. (\ref{diffusion})
diverges as $x^{-3/2}$, while the second term remains is zero
throughout the length of the contact.
Hence $f$ deviates from $f_T$ near the contact edges and the
squire-root singularity is smoothed out no matter how strong the e-e
scattering.

In this paper, we numerically calculate the high-frequency shot noise
for the simplest collision integral with energy-independent
transition probabilities:
\begin{eqnarray}
  I_{ee} (\epsilon) 
  = 
  -\frac{ \lambda_{ee} }{ \epsilon_{F} }
  \int  d\epsilon'
  \int  d\omega \,
& &
\nonumber\\
  \times
  \{ 
    f(\epsilon)
    f(\epsilon' - \omega) 
    \lbrack 
           1 
           - 
& &
           f(\epsilon - \omega) 
    \rbrack
    \lbrack 
           1 - f(\epsilon') 
    \rbrack 
\nonumber\\ 
    - 
    f(\epsilon - \omega) 
    f(\epsilon') 
    \lbrack 1 -
    f(\epsilon) 
    \rbrack 
& &
    \lbrack 
           1 - f(\epsilon' - \omega) 
    \rbrack
  \}.
  \label{3}
\end{eqnarray}
By doing so, we restrict ourselves to the Landau concept of
quasiparticle scattering and disregard the interference between e-e
and impurity scattering.\cite{AA85} This implies that the relevant
electron energies are sufficiently high: $\epsilon \gg \tau^{-1}
(p_F/k) (p_F l)^{-2}$, where $\tau$ is the elastic scattering time,
$p_F$ is the Fermi momentum, $k$ is the inverse Debye screening
length, and $l$ is the elastic mean free path of
electrons.\cite{AA79} In the gas approximation, where 
$k \ll p_F$, the dimensionless scattering amplitude equals
$\lambda_{ee} = \pi^2 k/64p_F$. However in realistic metals $k/p_F
\sim 1$, and $\lambda_{ee}$ should be renormalized by  
corrections of higher order in interaction. 

The actual behavior of $f$ near the contact edges may be understood
from the following semiquantitative reasoning. Select a point $x_0
\ll L$ near the left edge of the contact. Suppose that $f(x_0) = f_T$
with $T_e(x_0)$ given by Eq. (\ref{root}) and solve Eq.
(\ref{diffusion}) with boundary conditions $f(\epsilon, 0) =
\theta(-\epsilon)$ and $f(\epsilon, x_0) = f_T(\epsilon, x_0)$ in the
range $0 < x < x_0$. Because of the divergence of $\partial^2 f_T /
\partial x^2$ at $x = 0$ it is reasonable to expect that the
diffusion term in Eq. (\ref{diffusion}) will dominate over $I_{ee}$
at sufficiently small $x_0$ and the latter may be omitted. Then the 
resulting diffusion equation is easily solved and retaining only
terms linear in $x$, one obtains for the noise temperature $T_N(x) = 
(2\ln 2)\,T_e(x_0)x/x_0$. The crossover point $x_0$ is determined
from the condition that the diffusion term in Eq. (\ref{diffusion})
be of the order of the collision integral, i.e. $D/x_0^2 \sim
\lambda_{ee} T_e^2(x_0)/\epsilon_F$. This results in an estimate
$x_0 \sim L \alpha^{-1/3}$, where $\alpha = \lambda_{ee} (eVL)^2 / 
\epsilon_F D$ is the dimensionless parameter characterizing the 
relative strength of e-e interaction in the contact. From Eq. 
(\ref{general}), it 
follows that the limiting value of high-frequency noise is
\begin{equation}
  S_I(\infty)
  =
  \frac{ 2L }{ R }
  \left.
       \frac{ dT_N }{ dx }
  \right|_{ x = 0 }
  \sim
  eI
  \alpha^{1/6}.
  \label{limiting}
\end{equation}

The saturation frequency may be determined from the condition
$\gamma_{\omega} \sim x_0^{-1}$, which gives
$$
  \omega_s
  \sim
  \frac{ \sigma }{ \varepsilon_d }
  \frac{ \delta_0 r_0 }{ L^2 }
  \alpha^{2/3}.
$$

To test these semiqualitative conclusions, Eq.\ (\ref{diffusion}) was
numerically solved for different values of the dimensionless
parameter $\alpha$ using the finite-difference method on a lattice of
$100 \times 100$ sites. Figure \ref{temperature} shows the coordinate 
dependence of the noise temperature $T_N$ calculated for $\alpha = 
10$. It is clearly seen that in the middle of the contact, $T_N$ is 
close to $T_e$, while it remains almost unperturbed by the e-e 
interaction near the edges.

\begin{figure}
\epsfxsize18cm
\centerline{\epsffile{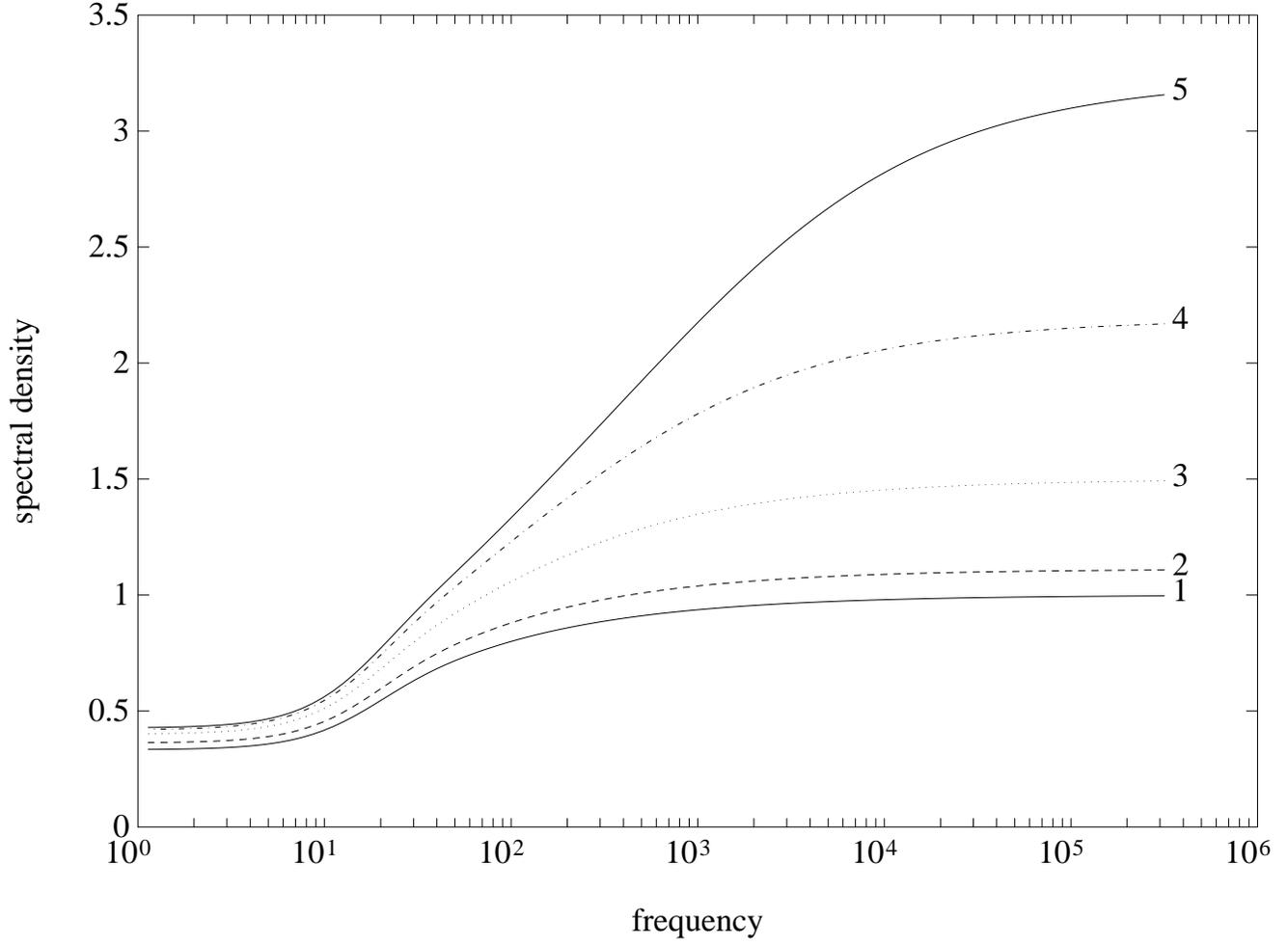}}
\vspace{0.3cm}
\caption
{
Dependences of normalized spectral density $S_I/2eI$ on dimensionless
frequency $\omega\varepsilon_d /4\pi\sigma\delta_0 r_0$ for (1)
$\alpha = 0$, (2) $\alpha = 10$, (3) $\alpha = 100$, (4) $\alpha =
1000$, and (5) $\alpha = 10000$.
}
\label{frequency}
\end{figure}

\begin{figure}
\epsfxsize18cm
\centerline{\epsffile{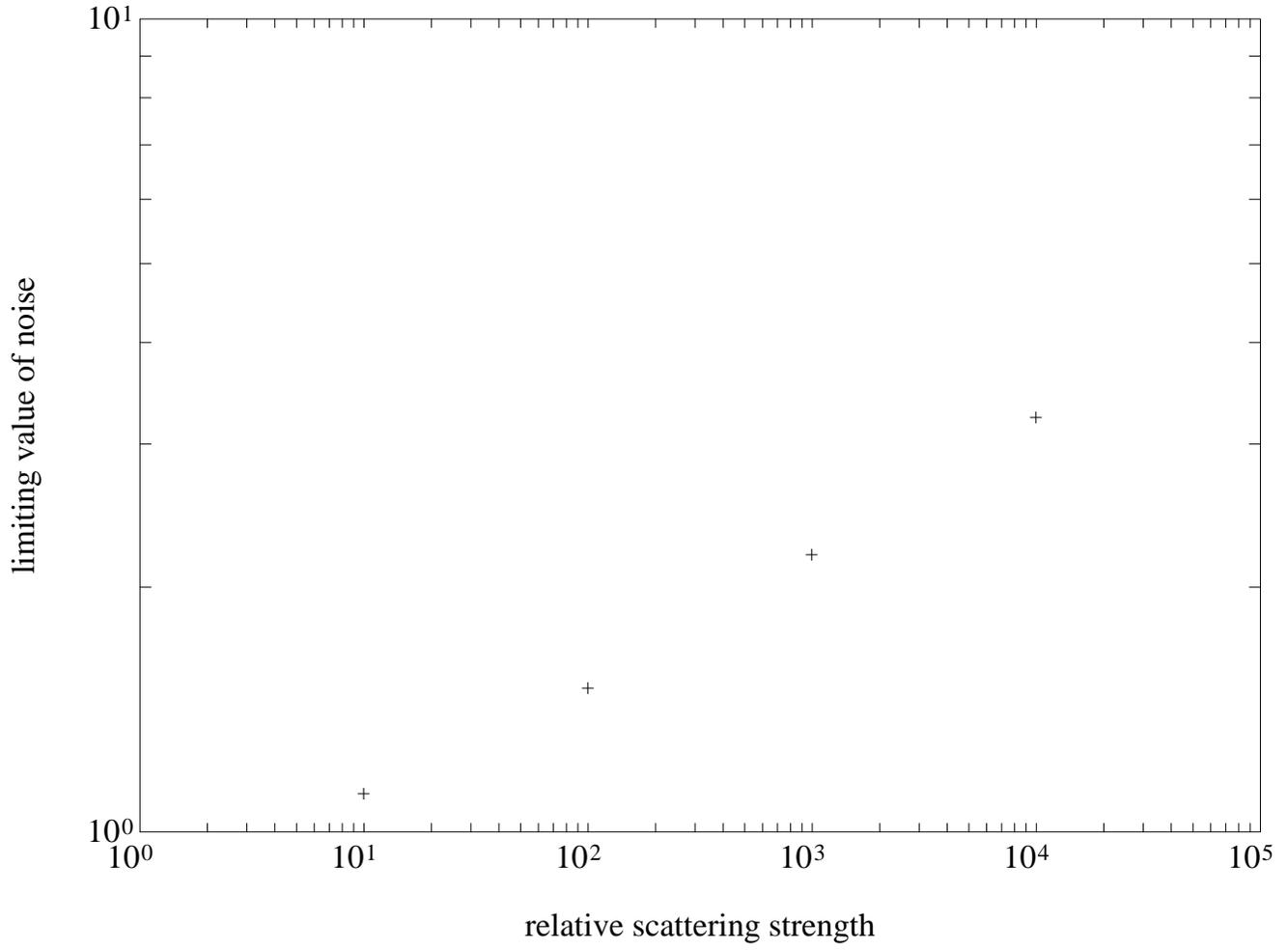}}
\vspace{0.3cm}
\caption
{
Log-log plot of saturated high-frequency noise $S_I(\infty)/2eI$ vs.
electron-electron scattering parameter $\alpha$.
}
\label{log-log}
\end{figure}

Figure \ref{frequency} shows the frequency dependences of noise for
five different values of $\alpha$ ranging from 0 to $10^4$. At zero
frequency, all the values of noise are located in a narrow range
$0.33 < S_I/2eI < 0.43$. As the frequency increases, the lower bound
for the noise and (especially) the spacing between the curves also
increase. In particular, this implies that at finite frequencies the
ratio $S_I/2eI$ is essentially voltage-dependent. The noise reaches
saturation for all $\alpha$ considered, but the saturation frequency
increases with $\alpha$. The saturation noise is $2eI$ for $\alpha =
0$ and also increases with $\alpha$. Note that for the maximum
of considered values $\alpha = 10^4$, the zero-frequency shot noise
coincides with the result of effective-temperature model to the third
decimal place, whereas the high-frequency noise is only nearly three
times the Poissonian one. The limiting values of high-frequency noise
are plotted versus $\alpha$ in Fig.\ \ref{log-log}. On a log-log
scale, this dependence presents almost a straight line except for the
lowest value $\alpha = 10$. At large $\alpha$, its slope corresponds 
to $S_I(\infty)/2eI \propto \alpha^{0.1683}$. This exponent is very
close to the value 1/6 that results from the above semiquantitative
reasoning. Possibly, the discrepancy could be made even smaller by
increasing the number of sites in the lattice and/or $\alpha$. In any
event, the above qualitative consideration provides a reasonably good
understanding of the behavior of high-frequency noise.

Naveh {\it et al.}\cite{Y98} proposed that the increase of
nonequilibrium noise at high frequencies may be used for
distinguishing between the cases of strong and weak e-e scattering.
Our calculations provide a basis for quantitative estimates of the
parameters of e-e scattering.  By measuring the voltage dependence of
saturated high-frequency noise and determining the corresponding
exponent, one may test the validity of Landau theory for this case.
Knowing the diffusion coefficient and $\epsilon_F$, it is also 
possible to determine the parameter of e-e interaction
$\lambda_{ee}$ from these data. In these measurements, the voltage
must be sufficiently high to exclude the effects of interference
between e-e and impurity scattering and to avoid quantum noise
dominating over the shot one.\cite{high-freq} Because of slow growth
of the noise-to-current ratio with voltage, proper care should be
taken to eliminate heating effects.

In summary, we have shown that in the presence of e-e scattering, the
shot noise in long diffusive contacts increases with frequency and
tends to a finite value. This value is larger than $2eI$ and
increases with voltage nearly according to the law $S_I(\infty)/2eI
\propto V^{1/3}$.

This work was supported by the Russian Foundation for Basic Research
(Project No. 96-02-16663-a).

\end{multicols}
\widetext
\newpage
\newpage
\newpage
\end{document}